%% file: main.tex
\documentclass[sigconf]{acmart}
\usepackage{multirow}
\usepackage{longtable}
\usepackage{colortbl}
\usepackage{soul}

\AtBeginDocument{%
  }

\setcopyright{acmlicensed}
\copyrightyear{2025}
\acmYear{2025}
\setcopyright{cc}
\setcctype{by-nc-sa}
\acmConference[CHI '25]{CHI Conference on Human Factors in Computing Systems}{April 26-May 1, 2025}{Yokohama, Japan}
\acmBooktitle{CHI Conference on Human Factors in Computing Systems (CHI '25), April 26-May 1, 2025, Yokohama, Japan}
\acmDOI{10.1145/3706598.3713780}
\acmISBN{979-8-4007-1394-1/25/04}

\begin{document}

\title[Implications of Indirect Speech in Physical Human-Robot Collaboration]{Can you pass that tool?: Implications of Indirect Speech in Physical Human-Robot Collaboration
}

\author{Yan Zhang}
\affiliation{%
  \department{School of Computing and Information Systems}
  \institution{University of Melbourne}
  \city{Melbourne}
  \state{VIC}
  \country{Australia}}
\email{yan.zhang.1@unimelb.edu.au}
\orcid{0000-0003-2142-5094}

\author{Tharaka Sachintha Ratnayake}
\affiliation{%
  \institution{University of Melbourne}
  \city{Melbourne}
  \state{VIC}
  \country{Australia}}
\email{tsratnayakem@student.unimelb.edu.au}
\orcid{0009-0004-6408-7587}

\author{Cherie Sew}
\affiliation{%
  \department{School of Computing and Information Systems}
  \institution{University of Melbourne}
  \city{Melbourne}
  \state{VIC}
  \country{Australia}}
\email{csew@student.unimelb.edu.au}
\orcid{0000-0002-5318-5681}

\author{Jarrod Knibbe}
\affiliation{%
  \department{School of Electrical Engineering and Computer Science}
  \institution{The University of Queensland}
  \city{Brisbane}
  \state{QLD}
  \country{Australia}}
\email{j.knibbe@uq.edu.au}
\orcid{0000-0002-8844-8576}

\author{Jorge Goncalves}
\affiliation{%
  \department{School of Computing and Information Systems}
  \institution{University of Melbourne}
  \city{Melbourne}
  \state{VIC}
  \country{Australia}}
\email{jorge.goncalves@unimelb.edu.au}
\orcid{0000-0002-0117-0322}

\author{Wafa Johal}
\affiliation{%
  \department{School of Computing and \\Information Systems}
  \institution{University of Melbourne}
  \city{Melbourne}
  \state{VIC}
  \country{Australia}}
\email{wafa.johal@unimelb.edu.au}
\orcid{0000-0001-9118-0454}

\renewcommand{\shortauthors}{Zhang et al.}

\begin{abstract}
Indirect speech acts (ISAs) are a natural pragmatic feature of human communication, allowing requests to be conveyed implicitly while maintaining subtlety and flexibility. Although advancements in speech recognition have enabled natural language interactions with robots through direct, explicit commands—providing clarity in communication—the rise of large language models presents the potential for robots to interpret ISAs. However, empirical evidence on the effects of ISAs on human-robot collaboration (HRC) remains limited. To address this, we conducted a Wizard-of-Oz study (N=36), engaging a participant and a robot in collaborative physical tasks. Our findings indicate that robots capable of understanding ISAs significantly improve human's perceived robot anthropomorphism, team performance, and trust. However, the effectiveness of ISAs is task- and context-dependent, thus requiring careful use. These results highlight the importance of appropriately integrating direct and indirect requests in HRC to enhance collaborative experiences and task performance.
\end{abstract}



\begin{teaserfigure}
  \begin{center}
  \includegraphics[width=0.85\textwidth]{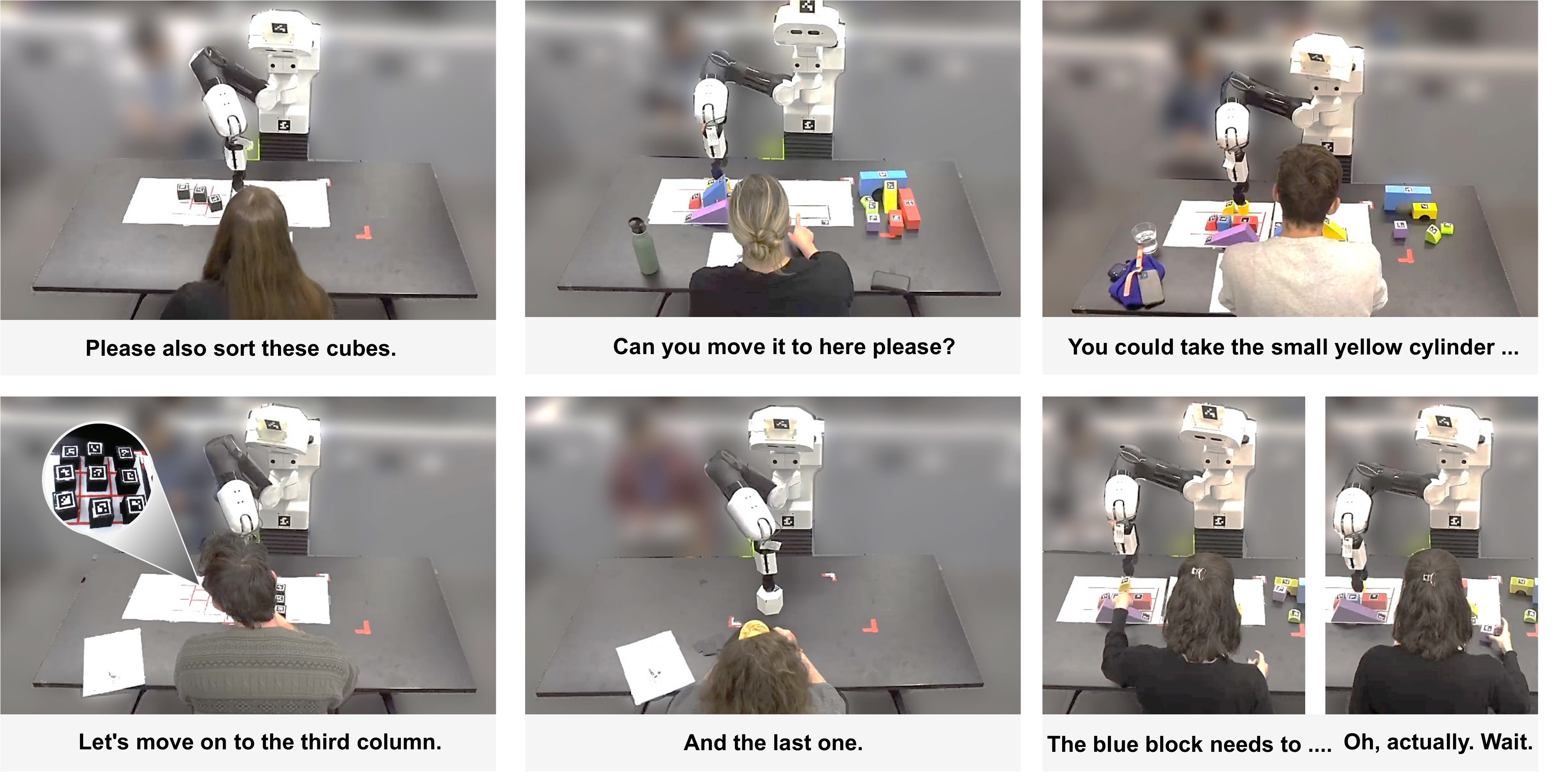}
  \end{center}
  \caption{This figure presents images from our experiment, featuring representative participant utterances to illustrate the types of requests used. The top left image depicts a direct request, while the rest of the images showcase various indirect requests. The interpretation and robot's responses are explained in \autoref{user_study}. }
  \label{fig:teaser}
\end{teaserfigure}


\maketitle

\input{sections/10-introduction}
\input{sections/20-related-works}
\input{sections/30-method}

\input{sections/40-result}
\input{sections/60-discussion}
\input{sections/70-conclusion}

\begin{acks}
We gratefully acknowledge the support provided by the Melbourne Research Scholarship (University of Melbourne) and the Australian Research Council Discovery Early Career Research Award (Grant No. DE210100858).
\end{acks}

\bibliographystyle{ACM-Reference-Format}
\bibliography{reference}

\appendix
\onecolumn
\section*{Appendix}
\renewcommand{\thetable}{A\arabic{table}}
\setcounter{table}{0} 

\section{Selected Questions from the Negative Attitude Towards Robots Scale} \label{apd:NARS}
``Please read through each of the following sentences and then indicate how frequently these statements apply to you, from (1) strongly disagree to (5) strongly agree.''
\begin{itemize}
    \item I would feel very nervous just standing in front of a robot.
    \item I would feel very nervous talking with robots.
    \item I would feel uneasy if I was given a job where I had to use robots.
\end{itemize}

\section{Model Results} \label{apd:model}

Table \ref{tb:model_result} provides the detailed results of the quantitative analysis.

The formula we used for the CLMM model in R is:
\[
dependent\_variable \sim speech\_mode + pre\_robot + pre\_va + pre\_phy + (1|task\_type) + (1|scale\_item) + (1+task\_type|p\_id)
\]
where dependent\_variable represents the rating from one of the scales (team fluency, goal alignment, performance trust, and anthropomorphism); speech\_mode represents the independent variable; pre\_robot, pre\_va, and pre\_phy are covariates, representing participants' previous experience with robots, voice assistants, and physical collaborative tasks; task\_type, scale\_item, and p\_id are random effects, representing tasks types, scales' sub-item IDs, and participant IDs.

\begin{table}[b]
\caption{This table shows the results from CLMM analysis. In the column of fixed effects, italics are covariates.}
\label{tb:model_result}
\begin{tabular}{|llllll|}
\hline
\multicolumn{6}{|c|}{\cellcolor[HTML]{D3DCD2}\textbf{Team Fluency}} \\ \hline
\textbf{Fixed Effects} & \textbf{Estimates} & \textbf{Std Error} & \textbf{95\% CI} & \textbf{z} & \textbf{p-value} \\ \hline
Speech Mode (Non-ISA) & 0.961 & 0.403 & 0.17 -- 1.751 & 2.382 & 0.017* \\
\textit{Robot (No)} & 0.004 & 0.110 & -0.211 -- 0.218 & 0.032 & 0.974 \\
\textit{Voice Assistant (Never)} & 0.129 & 0.210 & -0.283 -- 0.541 & 0.615 & 0.538 \\
\textit{Physical Collaborative Tasks (Never)} & 0.193 & 0.174 & -0.147 -- 0.533 & 1.112 & 0.266 \\ \hline
\textbf{Random Effects} & \textbf{Variance} & \textbf{Std Dev} & \textbf{Correlation} &  &  \\ \hline
Task Type & 0.097 & 0.311 &  &  &  \\
Scales' Sub-item ID & 0.072 & 0.269 &  &  &  \\
Participant ID & 1.528 & 1.236 &  &  &  \\
Task Type | Participant ID & 0.262 & 0.511 & -0.597 &  &  \\ \hline
\textbf{Model Fit} & \textbf{AIC} & \textbf{Log Lik} &  &  &  \\ \hline
 & 1009.02 & -489.51 &  &  &  \\ \hline
\multicolumn{6}{|c|}{\cellcolor[HTML]{D3DCD2}\textbf{Goal Alignment}} \\ \hline
\textbf{Fixed Effects} & \textbf{Estimates} & \textbf{Std Error} & \textbf{95\% CI} & \textbf{z} & \textbf{p-value} \\ \hline
Speech Mode (Non-ISA) & 2.309 & 0.656 & 1.023 -- 3.596 & 3.518 & \textless{}0.001*** \\
\textit{Robot (No)} & 0.120 & 0.170 & -0.214 -- 0.453 & 0.701 & 0.483 \\
\textit{Voice Assistant (Never)} & -0.099 & 0.316 & -0.719 -- 0.521 & -0.312 & 0.755 \\
\textit{Physical Collaborative Tasks (Never)} & 0.536 & 0.270 & 0.007 -- 1.064 & 1.985 & 0.047* \\ \hline
\textbf{Random Effects} & \textbf{Variance} & \textbf{Std Dev} & \textbf{Correlation} &  &  \\ \hline
Task Type & 0.050 & 0.224 &  &  &  \\
Scales' Sub-item ID & 0.000 & 0.000 &  &  &  \\
Participant ID & 4.266 & 2.065 &  &  &  \\
Task Type | Participant ID & 0.299 & 0.546 & -0.553 &  &  \\ \hline
\textbf{Model Fit} & \textbf{AIC} & \textbf{Log Lik} &  &  &  \\ \hline
 & 801.36 & -385.68 &  &  &  \\ \hline
\multicolumn{6}{|c|}{\cellcolor[HTML]{D3DCD2}\textbf{Performance Trust}} \\ \hline
\textbf{Fixed Effects} & \textbf{Estimates} & \textbf{Std Error} & \textbf{95\% CI} & \textbf{z} & \textbf{p-value} \\ \hline
Speech Mode (Non-ISA) & 1.105 & 0.493 & 0.138 -- 2.072 & 2.240 & 0.025* \\
\textit{Robot (No)} & -0.041 & 0.136 & -0.307 -- 0.226 & -0.298 & 0.766 \\
\textit{Voice Assistant (Never)} & 0.231 & 0.248 & -0.255 -- 0.717 & 0.932 & 0.351 \\
\textit{Physical Collaborative Tasks (Never)} & 0.400 & 0.211 & -0.014 -- 0.814 & 1.892 & 0.058 \\ \hline
\textbf{Random Effects} & \textbf{Variance} & \textbf{Std Dev} & \textbf{Correlation} &  &  \\ \hline
Task Type & 0.015 & 0.124 &  &  &  \\
Scales' Sub-item ID & 0.225 & 0.475 &  &  &  \\
Participant ID & 2.630 & 1.622 &  &  &  \\
Task Type | Participant ID & 0.019 & 0.136 & -1.000 &  &  \\ \hline
\textbf{Model Fit} & \textbf{AIC} & \textbf{Log Lik} &  &  &  \\ \hline
 & 990.83 & -482.42 &  &  &  \\ \hline

\multicolumn{6}{|c|}{\cellcolor[HTML]{D3DCD2}\textbf{Anthropomorphism}} \\ \hline
\textbf{Fixed Effects} & \textbf{Estimates} & \textbf{Std Error} & \textbf{95\% CI} & \textbf{z} & \textbf{p-value} \\ \hline
Speech Mode (Non-ISA) & 2.708 & 0.674 & 1.387 -- 4.03 & 4.016 & \textless{}0.001*** \\
\textit{Robot (No)} & -0.168 & 0.184 & -0.528 -- 0.192 & -0.915 & 0.360 \\
\textit{Voice Assistant (Never)} & 0.031 & 0.340 & -0.635 -- 0.697 & 0.092 & 0.927 \\
\textit{Physical Collaborative Tasks (Never)} & -0.111 & 0.288 & -0.676 -- 0.454 & -0.385 & 0.701 \\ \hline
\textbf{Random Effects} & \textbf{Variance} & \textbf{Std Dev} & \textbf{Correlation} &  &  \\ \hline
Task Type & 0.000 & 0.000 &  &  &  \\
Scales' Sub-item ID & 0.101 & 0.318 &  &  &  \\
Participant ID & 5.092 & 2.257 &  &  &  \\
Task Type | Participant ID & 0.832 & 0.912 & -0.552 &  &  \\ \hline
\textbf{Model Fit} & \textbf{AIC} & \textbf{Log Lik} &  &  &  \\ \hline
 & 1239.07 & -606.54 &  &  &  \\ \hline
\end{tabular}
\end{table}

\end{document}

%% file: sections/10-introduction.tex
\section{Introduction}
A spoken sentence is often not limited to its literal meaning. The question \textit{``Can you pass the salt?''} implicitly requests an action, while literally questioning the listener's physical ability to handover the salt. Alternatively, \textit{``This soup needs salt''} both asserts an opinion about the soup and, depending on the context and surrounding objects, may be requesting someone to pass the salt~\cite{clark1979responding}. These are examples of indirect speech acts (ISAs), which~\citet{searle1975indirect} defined as utterances where one speech act is performed indirectly by carrying out another, transforming direct intents into implicatures. They are complex, multi-faceted, and require shared context and interpretation~\cite{searle1975indirect}. They are also an optimized way to communicate that commonly occurs in collaboration settings where teammates build a shared understanding of the task~\cite{clark1986referring}. Similar to human-human interaction, understanding ISA in human-robot interaction is crucial, since interactions are often based on language to achieve a certain task or goal. 

The inherent naturalness, ease of production, and flexibility of indirect speech make it well-suited for effective human-robot collaboration (HRC)~\cite{obaigbena2024ai, liu2019review}. Through this lens, the robot is envisioned as a social, intelligent collaborator, where politeness, social etiquette, and discussion become factors of shared tasks. In emerging social collaborative robotic (cobotic) scenarios, such as with personal assistants in healthcare and accessibility~\cite{carros2020exploring, zhang2023follower}, then, ISAs are seen as a suitable method of interaction~\cite{williams2020excuse}. 

In performance- and task-oriented settings, however, the appropriateness of ISAs is less obvious. If the cobot partner is performing critical tasks, there may be little room for interpretation or lack of clarity. Direct speech -- \textit{``pass the salt''} -- is clear and timely. This is akin to more traditional interactions, where robots were clearly subservient and commands needed to be learned and delivered correctly. However, this learning creates barriers to natural and intuitive interaction, and the influence on user experience lacks evidence from comparative user studies. 
 
Even though the recent advances in large language models (LLMs) enhance the potential for natural speech interactions with robots in physical tasks~\cite{singh2023progprompt, macdonald2024language, liang2023code, zhao2024large}, to date, much of the attention is still on direct, explicit commands. This often oversimplifies communication, stripping away the naturalness observed in genuine human collaboration~\cite{lamm2017pragmatics}. The rate of LLMs' advances makes it likely that indirect speech will be supported through these models, before which, however, it remains essential to understand the role and impact of ISAs in human-robot collaboration.

Previous work has shown that humans tend to use ISAs when interacting with robots at frequencies similar to those used with other humans~\cite{lee2010receptionist, bennett2017differences}. This highlights the need to develop human-centred verbal communication interfaces for cobots that can accommodate the naturalistic and varied ways in which people express themselves. However, despite the indispensability of ISAs in collaborative communication, there remains a gap in empirical evidence regarding the impact of ISAs on human-robot collaboration, especially in tabletop manipulation tasks. 

To address this gap, we conducted a study with 36 participants, comparing two speech modes of a real robot in a laboratory setting: one capable of understanding ISAs and another without this capability, on three collaborative tasks. Given that natural language communication is a barrier preventing human-robot teams from outperforming human-human teams~\cite{schelble2023investigating}, we theorise that the use of ISAs can contribute to the effectiveness and naturalness of communication, thereby improving perceived team performance and user experience. Specifically, to assess the impact of ISAs on collaboration and communication, we evaluated four key metrics commonly used in HRC. \textit{Team fluency} reflects seamless coordination, which is critical for user satisfaction and acceptance of cobots~\cite{hoffman2019evaluating, duan2021bridging}. \textit{Goal alignment} measures the success and efficiency of collaboration~\cite{salehzadeh2022purposeful, salas1995situation}. \textit{Trust} is essential for preventing misuse or disuse of the robot, ultimately enhancing collaboration effectiveness~\cite{abbass2018foundations, zhang2023investigating}. Moreover, enabling the robot's ability to understand ISAs could serve as a means to induce \textit{anthropomorphism}, thereby improving collaborative engagement by fostering a sense of partnership and enhancing the collaborative experience~\cite{zhang2021ideal, rezwana2022understanding}.

In summary, our research addresses the following questions:
\begin{description}
    \item [RQ1] How does a robot's capability to understand indirect speech acts influence the perceived \textit{team's performance}?
    \begin{description}
        \item [RQ1.1] How does a robot's capability to understand indirect speech acts influence the \textit{fluency} of human-robot teamwork?
        \item [RQ1.2] How does a robot's capability to understand indirect speech acts influence the establishment of \textit{goal alignment} among the human-robot team?
    \end{description}
    \item [RQ2] How does a robot's capability to understand indirect speech acts influence a human teammate's \textit{trust} in the robot's performance?
    \item [RQ3] How does a robot's capability to understand indirect speech acts influence a human teammate's perception of the robot's \textit{anthropomorphism}?
\end{description}

Our findings show that while ISAs are beneficial in human-robot collaboration, their effectiveness can vary depending on the context. The quantitative results show the robot's ability to comprehend ISAs significantly enhances participants' perceived team performance, trust, and anthropomorphism. The use of ISAs fosters a deeper cognitive engagement, making the robot appear more as a collaborative partner rather than a mere tool. However, qualitative results suggest that the usage of ISA can be task- and context-dependent in human-robot collaboration, with inappropriate use potentially leading to negative impacts on trust and user perception. These insights highlight the inherent limitations of relying solely on direct command-based interactions, which lack the subtlety required for establishing shared understanding and the sense of teaming. They also emphasise the importance of using indirect requests in a contextually adaptive and appropriate manner. Therefore, the careful integration of direct and indirect verbal communication emerges as a critical factor in optimising the performance and overall experience of human-robot collaboration. We advocate for the human-computer interaction (HCI) and human-robot interaction (HRI) community to develop human-centred LLMs for collaborative robots, recognising the critical role of ISAs in achieving this goal.

%% file: sections/20-related-works.tex
\section{Related Work}
\subsection{Speaking to Embodied Agents}
Voice assistants (VA), like Siri and Alexa, can have a noticeable influence on user behaviour, as these voice command interfaces are increasingly integrated into daily interactions through devices like phones, computers, and cars.~\cite{amershi2019guidelines, tsoli2018interactive}. The reach of VAs extends beyond simple task execution, influencing users' linguistic habits and potentially shaping social norms surrounding technology use~\cite{motta2021users, williams2020excuse}. Early research primarily addressed the technical challenges associated with speech detection and dialogue systems, focusing on improving the accuracy and efficiency of voice recognition technologies~\cite{cohen1995role}. As VAs became more commercialised, researchers observed that people adapt their language when interacting with VAs, using direct commands, simplified sentences, and keywords to mitigate the risk of misinterpretation~\cite{jaber2024cooking, myers2018patterns}. This adaptation reflects the users' low expectations of language processing and voice interfaces, as well as the inherent limitation of VAs' ability to comprehend and execute complex commands accurately. However, with the advancements in natural language processing, there has been a shift away from command-based paradigms towards more nuanced and complex verbal interactions due to its increased ability to infer intention and understand context~\cite{tanneberg2024help, ning2024user}, allowing for more natural and fluid dialogues between users and machines~\cite{abdul2015survey, yi2024survey}. 

Voice command interfaces have been implemented in embodied agents, such as social and collaborative robots, offering significant advantages in making these systems more human-like assistants capable of supporting real-world tasks. The advantages of incorporating voice interfaces into robots are evident, particularly in scenarios where natural and intuitive communication is essential. Besides, voice interfaces make AI and digital information more accessible to specific populations, such as children and the elderly, who may otherwise struggle with traditional interaction methods~\cite{shiomi2015effectiveness, ruggiero2022companion}. To make the voice interface more capable and better accommodate human activities, recent studies have increasingly focused on elements such as vocal fillers~\cite{ohshima2015conversational}, voice-matching~\cite{de2024your}, and social norms, particularly language politeness~\cite{williams2020excuse}. Among these, most elements contribute to making interactions feel more natural and human-like. Anthropomorphism remains one of the most extensively studied characteristics in human-agent verbal interaction~\cite{seaborn2021voice}. 

A considerable amount of existing research has explored the potential impact of robots' voices on human perception of human-likeness, trust, and capability. For instance, studies have shown that a human-like voice can increase trust in the robot~\cite{xu2019first}, with this effect being more noticeable when the robot's voice matches the gender of the participant~\cite{eyssel2012if}. Another study found that participants issued more commands to robots with artificial voices compared to those with human-like speech, suggesting that a less human-like voice may lead users to perceive the robot as a less capable machine rather than as a competent human~\cite{sims2009robots}. While the anthropomorphism of robots' speech can enhance user experience, it also increases the risk of participants overestimating the robot's intelligence and abilities~\cite{cha2015perceived}. Other factors like politeness, humor, and directness also shape a robot's perceived anthropomorphism ~\cite{emnett2024using}. Robots using indirect language in social interactions often seem more human-like ~\cite{saunderson2021robots}.

\subsection{Verbal Communication during Human-Robot Collaboration}
Verbal communication offers distinct advantages due to its naturalness and efficiency~\cite{liu2019review}. Previous research has demonstrated that humans communicating task-related information to robots can enhance the robot's understanding of goals and intentions, thereby improving overall performance~\cite{breazeal2004designing}. Additionally, studies have shown that robots equipped with communication abilities and verbal feedback can improve team performance by reducing task completion times and being perceived as better teammates~\cite{st2015robot}. Furthermore, explicitly incorporating context into communication enhances clarity, reduces ambiguity, and improves mutual understanding~\cite{mavridis2005grounded, tellex2014asking}.

In natural language processing for human-robot collaboration, several methods exist to parse commands from explicit utterances. The most direct approach involves extracting semantic features and mapping them to predefined robot controllers~\cite{tellex2006spatial}. However, research has shown that participants often provide instructions at varying levels of abstraction~\cite{anderson2018vision}. To interpret more abstract commands lacking specific keywords, association models are used to combine literal linguistic features and extract semantic meaning, typically relying on probability-based methods~\cite {misra2016tell, liu2016natural, matuszek2013learning}. Additionally, to generalize across new tasks and enable contextual understanding, researchers are exploring the usage of large language models for controlling robots in physical tasks~\cite{singh2023progprompt, macdonald2024language, liang2023code, zhao2024large}. While LLMs have the potential to comprehend implicit verbal commands, most studies focus on explicit, direct commands, which provide clear instructions but do not capture the nuanced and indirect nature of human communication in real-world scenarios.

However, relying primarily on direct commands that explicitly convey human requests oversimplifies interactions. Indirect speech acts are a natural feature of human communication, contributing to enhancing robots' anthropomorphism, which has been shown to be an important factor in creating an ideal AI teammate~\cite{zhang2021ideal}. ISAs also serve as an implicit and important means for humans to express their intentions. When the ISAs are misinterpreted during collaboration, the potential for long-term efficiency gains is compromised. Therefore, equipping robots with the ability to interpret ISAs enables them to respond more naturally and effectively, closely mimicking human-like communication patterns. This capability is particularly important in tasks that require high levels of coordination and mutual understanding, such as cooperative manipulation tasks~\cite{shah2011improved}.

\subsection{Indirect Speech Acts in Human-Robot Interaction}
Research has shown that humans tend to use indirect verbal requests when interacting with robots at frequencies similar to those used in human-human interactions, demonstrating the necessity of enabling ISAs in HRI~\cite{lee2010receptionist, bennett2017differences}. Several studies have focused on providing robots with the ability to interpret indirect requests. For example, Briggs and Scheutz~\cite{briggs2013hybrid} created a hybrid system to comprehend indirect requests and provide appropriate responses. Another studies~\cite{williams2015going, wen2020dempster} introduced a probabilistic algorithm for robots to learn sociocultural norms, infer intentions from human utterances, and generate clarification requests.

Moreover, the impact of ISAs on human-robot interaction has been examined from various perspectives. Research shows that robots employing ISAs are perceived as more likeable~\cite{torrey2013robot, strait2014let}, trustworthy~\cite{saunderson2021robots}, and willing to help~\cite{srinivasan2016help}. Conversely, a robot's inability to understand conventionalised ISAs (e.g., \textit{``Can you..?''}, \textit{``I need you to..''}) during social interactions has been found to negatively affect its performance and human perception~\cite{williams2018thank}. Even when participants are aware that robots may not fully comprehend ISAs, they tend to continue using them, which can impact the interaction fluency~\cite{briggs2017enabling}. While current research largely focuses on social interactions, there is still a limited exploration of ISAs in the context of physical collaboration, where conversations tend to be more collaborative, continuous, and shaped by physical context, rather than purely by politeness and social norms.

Although ISAs enable robots to engage in more nuanced and contextually rich social interactions, there is still a gap in the empirical evaluation of ISAs' impact on perceived task performance and user experience in HRC across physical collaborative tasks. In this study, we investigate the effects of a robot's ability to understand ISAs on team fluency, goal alignment, and human perception based on the task taxonomy for robotic manipulators concluded by~\cite{semeraro2023human}.

\subsection{Hypotheses}
Based on prior literature, we outline several hypotheses to address the research questions.

Previous research highlights that implicitly conveying contextual information through language can foster mutual understanding and facilitate smoother teamwork~\cite{frank2012predicting, clark1996using}. Therefore, we hypothesise that enabling the robot to understand ISAs will positively influence team fluency and goal alignment.
\begin{description}
    \item [H1a] Perceived team fluency will be better when the robot has the capability to understand ISAs. 
    \item [H1b] Perceived goal alignment will be better when the robot has the capability to understand ISAs.
\end{description}

Existing literature suggests that ISAs can increase trustworthiness in social interaction scenarios~\cite{saunderson2021robots}. However, there is a lack of research on their impact in physical collaborative scenarios. In this study, we hypothesise that a robot's ability to understand ISAs will positively impact trust in physical collaboration contexts.
\begin{description}
    \item [H2] Participants will perceive the robot as more trustworthy when it has the capability to understand ISAs.
\end{description}

Research shows that more human-like robots are perceived as better teammates~\cite{zhang2021ideal}. Moreover, human-like communication has been statistically proven to be highly effective in enhancing the impact of anthropomorphism compared to other anthropomorphic morphologies~\cite{roesler2021meta}. Thus, we hypothesise that the robot's ability to understand ISAs will enhance the user experience by increasing its perceived anthropomorphism, making interactions feel more human-like. 
\begin{description}
    \item [H3] Participants will perceive the robot as exhibiting greater anthropomorphism when it has the capability to understand ISAs.
\end{description}

%% file: sections/30-method.tex
\section{User Study} \label{user_study}
To investigate the impact of a robot's ability to understand indirect speech acts on people's perception, we conducted a Wizard-of-Oz experiment with 36 participants on three different physical collaborative tasks.

The experiment employed a mixed-method experimental design~\cite{creswell1999mixed}, collecting quantitative data through a questionnaire on team fluency, goal alignment, performance trust, and anthropomorphism as dependent variables, as well as qualitative data from interview responses. The Speech Mode (ISA vs. Non-ISA) served as a between-subject factor, with half of the participants interacting with a robot capable of understanding ISAs, while the other half interacted with a robot unable to comprehend ISAs. Each participant completed three tasks in counter-balanced order with the robot using one of the assigned Speech Modes, which was followed by a semi-structured interview. We provide additional detail on our experimental design in the following sections.

\begin{figure*}
    \centering
    \includegraphics[width=\textwidth]{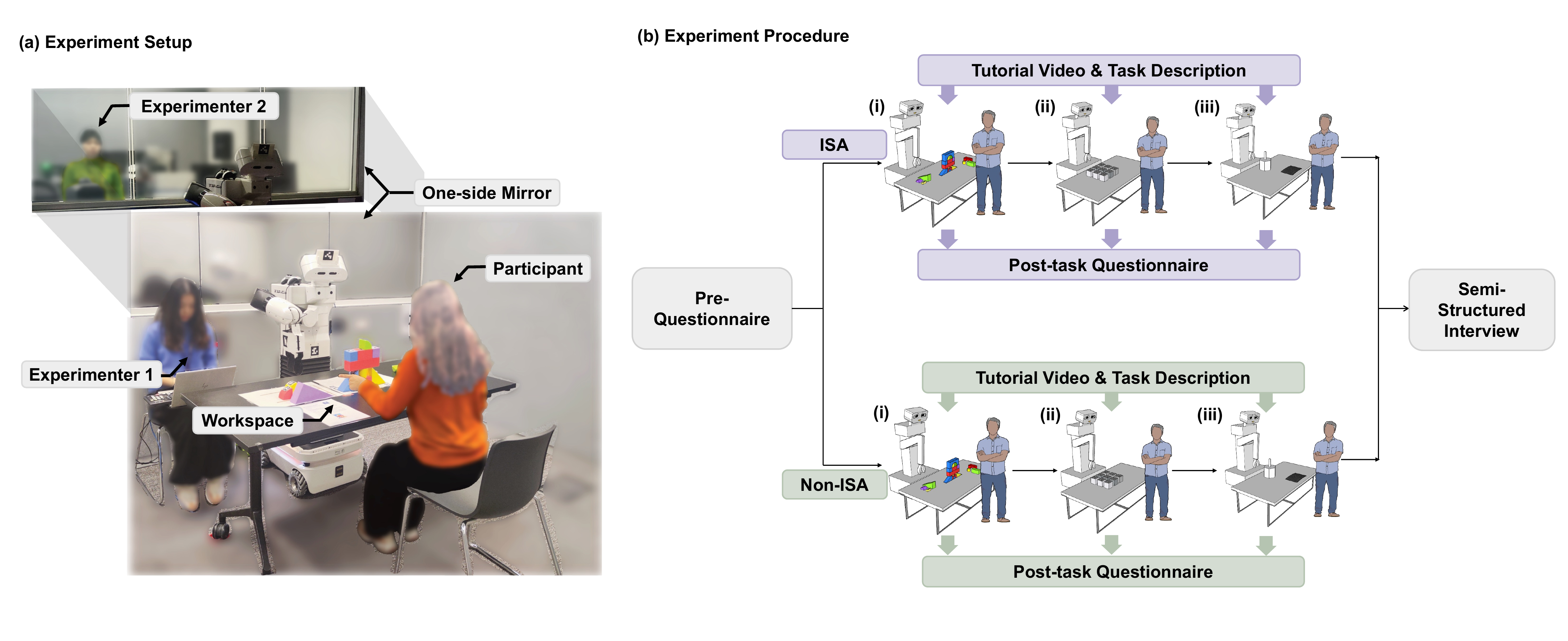}
    \caption{(a) Experiment setup: The participant, robot, and experimenter 1 were all present in the same room. The participant and robot were seated on opposite sides of a table, with the shared workspace located in the centre. Experimenter 1 (Speech Wizard) sat next to the robot, near the emergency button, and operated the robot's speech-WoZ interface. Experimenter 2 (Motion Wizard) was positioned behind a one-side mirror, allowing for a clear view of the room, and was responsible for teleoperating the robot's arm movements. (b) Experiment procedure: Each participant first completed a pre-questionnaire before being assigned to either the ISA or non-ISA group. The participant then performed three tasks with the robot in a counter-balanced order. Before each task, participants watched a tutorial video and read a task description. After completing each task, they filled out a post-task questionnaire. The experiment concluded with a semi-structured interview.\vspace{-1em}}
    \label{fig:exp}
\end{figure*}

\subsection{Experimental Design}
\subsubsection{Apparatus and Setup}
We used TIAGo as the robot agent in our study. TIAGo is a mobile manipulator robot with anthropomorphic features, including a head, neck, torso, and arm, making it well-suited for HRI research~\cite{pages2016tiago}. The robot and the participant were on the opposite side of a table, which acted as the shared workspace between the two parties. Given the current limitations of algorithms in achieving human-level understanding and generating accurate verbal responses to ISAs, and to minimise the influence of potential robotic failures on experimental outcomes, we chose to use a Wizard-of-Oz (WoZ) approach. WoZ is a classic methodology in HRI research, where human operators discretely control the robot’s behaviour to simulate advanced robotic capabilities that the system itself may not be able to achieve autonomously yet or that would not be robust for real-time interaction~\cite{martelaro2016wizard}. This approach allows researchers to focus on understanding user interactions with the robot without being hindered by technological limitations or safety issues related to autonomous motion control. 
In this experiment, two experimenters discreetly controlled the robot to provide realistic and fluid responses, enabling a better assessment of human-robot interaction dynamics.

One of the experimenters (Motion Wizard) was teleoperating the robot's movement behind a one-side mirror, which allowed them to have a clear view of both the robot and the participant while remaining hidden from the participant.
This teleoperation was possible thanks to custom-made software developed by our team that allowed the Motion Wizard to send commands to the robot remotely. This WoZ software was built using the Robotics Operating System (ROS) and the TIAGo API. We implemented the arm actions using inverse kinematics, which calculates the joint configuration based on the desired Cartesian coordinates of the end effector~\cite{chitta2017}. In addition to moving the end effector within a 3D space above the workspace, the robot's head also had 2 degrees of freedom, which allowed the Motion Wizard to observe through the robot's camera and actively engage with the participant. Safety was ensured by a collision detection function that automatically disabled the arm controller when abnormal tolerance values were detected in the joints. Virtual walls were also implemented around the robot's arm to restrict its movement, preventing it from exceeding a designated range or approaching the participant too closely.

The other experimenter (Speech Wizard) was sitting beside the robot's emergency button and operating the speech-WoZ interface through a laptop to give verbal responses, which were scripted in advance (See \autoref{speech_model} in detail). Participants were informed that the Speech Wizard served as a safeguard, responsible for ensuring their physical safety by using the emergency button located on the robot's base if necessary. This explanation led participants to view the Speech Wizard’s presence as a precautionary measure. TIAGo utilises Acapela Group's Text-to-Speech technology, which carries out the phonetic transcription of the text, generates prosody for the speech, produces the audio signal, and plays through TIAGo's speaker. \autoref{fig:exp}a demonstrates the experimental setup.

\begin{table*}[t]
\caption{Examples of participants' requests, interpretations, and robot's responses in different Speech Modes. The request examples are from \autoref{fig:teaser}. (P: participant; R: robot)}
\label{tb:example}
\resizebox{0.95\textwidth}{!}{
\begin{tabular}{llll}
\hline
\multirow{2}{*}{Request Examples} &
  \multirow{2}{*}{Interpretations} &
  \multicolumn{2}{c}{Robot's Responses} \\ \cline{3-4} 
 &
   &
  If in the ISA group &
  If in the Non-ISA group \\ \hline
P: Please also sort these cubes. &
  \begin{tabular}[c]{@{}l@{}}\textbf{Direct}\\ \textit{Literal}: Sort cubes.\\ \textit{Intent}: Sort cubes.\end{tabular} &
  \multicolumn{2}{c}{\begin{tabular}[c]{@{}c@{}}R: Yes, sure. (Act on the intent)\end{tabular}} \\ \hline
  \begin{tabular}[c]{@{}l@{}}P: Can you move it to here\\ please? \end{tabular} &
  \begin{tabular}[c]{@{}l@{}}\textbf{Indirect}\\ \textit{Literal}: Ask for the ability to move it.\\ \textit{Intent}: Move it.\end{tabular} &
  \begin{tabular}[c]{@{}l@{}}R: Got it.\\ (Act on the intent)\end{tabular} &
  \begin{tabular}[c]{@{}l@{}}R: Yes, I can do that.\\ (No action)\end{tabular} \\ \hline
\begin{tabular}[c]{@{}l@{}}P: You could take the small\\ yellow cylinder ...\end{tabular} &
  \begin{tabular}[c]{@{}l@{}}\textbf{Indirect}\\ \textit{Literal}: Suggest an action option to\\ take the cylinder.\\ \textit{Intent}: Take the cylinder.\end{tabular} &
  \begin{tabular}[c]{@{}l@{}}R: Okay.\\ (Act on the intent)\end{tabular} &
  \begin{tabular}[c]{@{}l@{}}R: Well noted.\\ (No action)\end{tabular} \\ \hline
 \begin{tabular}[c]{@{}l@{}}P: Let's move on to the third\\ column.  \end{tabular} &
  \begin{tabular}[c]{@{}l@{}}\textbf{Indirect}\\ \textit{Literal}: Suggest moving on to the third\\ column.\\ \textit{Intent}: Sort the third column.\end{tabular} &
  \begin{tabular}[c]{@{}l@{}}R: Working on that.\\ (Act on the intent)\end{tabular} &
  \begin{tabular}[c]{@{}l@{}}R: It's a good suggestion.\\ (No action)\end{tabular} \\ \hline
P: And the last one. &
  \begin{tabular}[c]{@{}l@{}}\textbf{Indirect}\\ \textit{Literal}: A reference to the last thing.\\ \textit{Intent}: Rotate to the last face.\end{tabular} &
  \begin{tabular}[c]{@{}l@{}}R: Sure.\\ (Act on the intent)\end{tabular} &
  \begin{tabular}[c]{@{}l@{}}R: ...\\ (Silence. No action)\end{tabular} \\ \hline
\begin{tabular}[c]{@{}l@{}}P: The blue block needs to ...\\ P: Oh, actually. Wait.\end{tabular} &
  \begin{tabular}[c]{@{}l@{}}\textbf{Indirect}\\ \textit{Literal}: Provide information for the blue\\block and wait.\\ \textit{Intent}: Move the blue block to a position. \\ The last command is wrong, stop the\\ current action and wait for the next one.\end{tabular} &
  \begin{tabular}[c]{@{}l@{}}R: Got it.\\ (Act on the intent, \\ then stop halfway)\end{tabular} &
  \begin{tabular}[c]{@{}l@{}}R: Thank you for the info.\\ (No action)\end{tabular} \\ \hline
\end{tabular}
}
\end{table*}

\subsubsection{Robot's Speech Understanding} \label{speech_model}
The robot's Speech Mode was a between-subject independent variable with two conditions. In the ISA condition, the robot could understand participants' ISAs and respond with appropriate actions. In the Non-ISA condition, the robot was only able to grasp the literal meaning of requests and respond to commands that were stated in imperative sentences. The literal meaning of ISAs was interpreted by isolating them from their contextual elements, following the guidelines of Searle's putative facts~\cite{searle1975indirect}. We selected some representative requests from participants to demonstrate how the direct and indirect speech acts were interpreted and responded to during our experiment (shown in \autoref{tb:example} and \autoref{fig:teaser}). To respond to both indirect and direct requests, the speech-WoZ interface featured predefined sentences, such as ``Sure,'' ``Okay, working on that,'' and ``Yes, I have the ability to do that.'' Utterances without command intent, such as ``Thank you, TIAGo,'' were responded to as natural conversational exchanges like ``You're welcome.'' The selection of phrases was guided by the aforementioned literature and further refined through insights gained from four pilot studies. The interface also provided a text box that allowed the Speech Wizard to input responses to any unexpected speech. To maintain the flow of interaction and avoid constraining the use of ISAs, participants were allowed to use gestures along with their speech, which experimenters interpreted and responded to accordingly. Notably, no participants reported noticing that the experimenter sitting in front of them was controlling the robot's speech.

\subsubsection{Collaboration Tasks}
A recent systematic review~\cite{semeraro2023human} categorised the HRC tasks for robotic manipulators as: (1) collaborative assembly, where humans and robots work together to assemble complex objects through a series of sequential sub-processes; (2) object handling \& handover, involving the joint grasping and placement of objects by humans and robots, as well as the handover of objects from the robot to the human; and (3) collaborative manufacturing, where both humans and robots perform tasks that permanently alter an object, such as polishing and drilling. For safety considerations, we modified the object handling and handover task to a turn-based pick-and-place activity, where both the robot and the human participated in sorting cubes. Based on this taxonomy, we designed and implemented three physical collaborative tasks for our experiment: (1) a foam brick assembly task~\cite{vogt2016learning}, (2) a 3*3 cubes sorting task~\cite{faroni2020layered}, and (3) a hexagonal prism polishing task~\cite{nikolaidis2015efficient}. 
In each task, the robot lacked prior information about the task's goal and plan, requiring the participant to relay the instructions to the robot at the beginning and verbally guide the team's actions throughout the entire activity.

The \textbf{assembly} task (\autoref{fig:exp}bi) required the human-robot team to build a structure using foam bricks. We distributed 18 bricks of various shapes between the human and robot, with 12 bricks required for constructing the target structure and 6 incorrect bricks that should not be used.  
Only the participant was provided with a photo of the structure they had to build, while the robot had no prior knowledge of the structure. The bricks were initially randomly placed in the robot or the participant's stock, and each party was only allowed to take bricks from their own pile. 
Participants could only manipulate the bricks on their side and needed to communicate and coordinate with the robot to have it add its bricks to the construction. 

The \textbf{sorting} task (\autoref{fig:exp}bii) used nine 5*5*5cm cubes that needed to be rearranged according to two categorical attributes: texture and version.  Each cube featured a type of surface texture (smooth, medium, rough), and an ArUco marker~\cite{garrido2014automatic} encoding its version information (old, intermediate, new). 
To mimic an information asymmetry sorting task, the participants were able to touch and feel the texture, whereas the robot could scan the ArUco marker to access the cube's version.
We used apparently similar ArUco markers for version information to make it impossible for participants to distinguish between them by sight alone. Only the exchange of information between teammates made it possible to achieve the task: to arrange the cubes in a gradient from rough to smooth in one dimension and from old to new in the orthogonal direction. 

The \textbf{polishing} task (\autoref{fig:exp}biii) constituted a simple instantiation of a manufacturing task. The robot was responsible for holding and turning the hexagonal prism, and the participant polished each surface three times using sandpaper. Every time the participants were happy with the sanding, they had to communicate to the robot to turn the object to show a face that had not been polished. This scenario was designed to simulate a situation where the hexagonal prism was too heavy or hazardous for a human to lift and rotate, requiring cooperation with the robot to successfully complete the task.

\subsection{Participants}
We conducted \textit{a priori} power analysis to calculate the sample size for our experiment using \textit{G*Power}~\cite{faul2007g}. The calculation was based on a medium effect size of $f=0.25$, an alpha-level of 0.05, and a power of 0.9. As a result, we recruited 36 ($Female:Male = 19:17$,  $M_{age} = 24.08$, $Std_{age} = 5.75$) participants who were all fluent English speakers. We used the three questions of the interaction subscale from the Negative Attitude Toward Robots Scale (NARS Questionnaire) \cite{nomura2006measurement}, as they were relevant to working and talking to a robot (see \autoref{apd:NARS}). These questions were used to screen out individuals who exhibited strong negative responses towards robots and who could possibly feel distressed interacting with a robot (i.e., a rating higher than 3). Each experiment took about 60 minutes and participants were compensated with a \$30 voucher. Our experiment received ethics approval from the Institutional Review Board (IRB).

\subsection{Procedure}
The experiment procedure is shown in \autoref{fig:exp}b. Upon welcoming the participants, the study started with a pre-questionnaire, which captured participant demographics and their prior interaction experience with robots, voice assistants, and in performing physical collaborative tasks. The prior experience served as covariates in data analysis. Before each task, participants were provided with a tutorial video and a written task description, which included instructions and specified the objectives of the task. Additionally, a picture of the target structure for the assembly task was presented to illustrate the final goal. 
Participants were required to lead the collaboration and verbally relay the team's objectives to their robot teammate, TIAGo. After each task, participants completed a post-task questionnaire that assessed their perceptions of the team's fluency and goal alignment~\cite{hoffman2010effects}, performance trustworthiness using Multi-Dimensional Measure of Trust (MDMT)~\cite{malle2021multidimensional}, and the robot's anthropomorphism using the Godspeed Questionnaire (GSQ)~\cite{bartneck2023godspeed}. Each participant interacted with one of the robot's Speech Modes (ISA or Non-ISA) and engaged in three tasks, which were assigned in a counter-balanced order. Finally, the study ended with a semi-structured interview. Each experiment took about 60 minutes, including the interview.

\subsection{Data Collection and Analysis}
We collected the quantitative data using standard questionnaires and the qualitative data through a semi-structured interview. The following dependent variables were collected after each task:

\begin{itemize}
    \item Team fluency: To answer RQ1.1, we used the 7-point team fluency sub-scale with 3 items, from~\cite{hoffman2010effects}, which adapted the Working Alliance Inventory~\cite{horvath1989development} on Human-Robot Collaboration.
    \item Goal alignment: For the goal alignment in RQ1.2, we utilised the 7-point goal sub-scale with 3 items, from~\cite{hoffman2010effects}. 
    \item Performance trust: The 4-item MDMT performance trust scale results were collected to measure participants' perceived capability and reliability of the robot (RQ2). The scale has 5 points and an additional option for ``Does not fit'' to prevent forced and possibly meaningless ratings~\cite{malle2021multidimensional}.
    \item Anthropomorphism: To answer RQ3, the 5-point anthropomorphism sub-scale with 5 items of the GSQ was used. As the study was focused on the robot's understanding of communication rather than the appearance of the robot, the last item, ``Moving rigidly/elegantly'', was changed to ``communicating rigidly/elegantly'', which has been shown to be reliable by~\cite{laban2019working}. 
\end{itemize}

\vspace{-2em}

To analyse the impact of the robot's Speech Modes (ISA vs. Non-ISA) and covariates (participants' prior interaction experience with robots, voice assistants, and physical collaborative tasks), we used Cumulative Link Mixed Models (CLMMs) via the "ordinal" package in R~\cite{christensen2019ordinal}. This analysis is appropriate given the ordinal nature of our dependent variables. Additionally, task type, scales' sub-item ID, and participant ID were included as random effects in our model to account for potential variability within group structures and repeated measures~\cite{brown2021introduction}.

At the end of the experiment, we conducted a semi-structured interview lasting approximately 15 minutes to gather qualitative feedback from participants. The Motion Wizard observed participants' behaviours during the experiment. Instances of participants using indirect speech acts were further explored through follow-up questions during the interviews. The interviews were intended to supplement the quantitative results and provide insight into their subjective feelings regarding the overall experience during the collaboration.
Given the between-subjects design of the study, we began the interview by explaining the experimental condition that participants had not experienced, ensuring they had a comprehensive understanding of the study. We disclosed that the experimenters controlled the robot's actions and speech only after the interview concluded.

The interview results were transcribed and analysed through reflexive thematic analysis (RTA), which was well-suited to this study because it emphasised the researchers’ active role in constructing themes, thereby fostering flexibility, creativity, and critical reflection. This approach permits researchers to integrate their own insights and observations from the experimental process, making it particularly effective for exploring subtle phenomena~\cite{braun2023doing}. Following the 6-phase guidance by~\cite{braun2006using}, two authors of this paper, both of whom possess substantial expertise in human-robot interaction and human-computer interaction, conducted the RTA. In phase 1, researchers thoroughly reviewed all transcriptions. In phase 2, they inductively generated initial codes at the sentence level, which were either semantic, representing participants' explicit feelings, or latent, reflecting deeper meanings inferred from the data based on researchers' knowledge background. In phase 3, they constructed the initial themes and categorised the codes. Up to this point, the work had been carried out individually by each researcher. In phase 4, two researchers cooperatively discussed and reviewed the themes through multiple rounds. In phase 5, the themes were defined and named. In phase 6, researchers drafted the initial report of the qualitative analysis. Phases 4 to 6 were repeated over several rounds, during which the themes were iteratively refined and discrepancies addressed. This process aligns with the RTA principles, which emphasise continuous iterative reflexivity to ensure the analysis remains progressively recursive~\cite{terry2017thematic}.

%% file: sections/40-result.tex
\section{Results}
\autoref{tb:demo} shows a summary of participants' demographics and their prior interaction experience with robots, voice assistants, and physical collaborative tasks (with either humans or robots). Next, we report our quantitative and qualitative findings. 

\begin{table*}[]
\caption{Overview of participants ' demographic information and their prior interaction experience with robots, voice assistants, and physical collaborative tasks. (Rarely: less than once a month; Sometimes: at least once a month but less than once a week; Often: at least once a week but less than once a day; Very often: at least once a day.)}
\label{tb:demo}
\begin{tabular}{lclclclclc}
\hline
\multicolumn{2}{c}{Gender} & \multicolumn{2}{c}{Age} & \multicolumn{2}{c}{Robots} & \multicolumn{2}{c}{Voice Assistant} & \multicolumn{2}{c}{Physical Collaborative Tasks} \\ \hline
Female & 52.8\% & 18-25 & 72.2\% & No experience           & 33.3\% & Never      & 5.6\%  & Never      & 25.0\% \\
Male   & 47.2\% & 26-35 & 19.4\% & With domestic robot     & 55.6\% & Rarely     & 55.6\% & Rarely     & 22.2\% \\
       &        & 36-45 & 8.3\%  & With desktop pet robot   & 2.8\%  & Sometimes & 19.4\% & Sometimes & 27.8\% \\
       &        &       &        & With social robot       & 0.0\%  & Often      & 13.9\% & Often      & 19.4\% \\
       &        &       &        & With industrial robot    & 8.3\%  & Very often & 5.6\%  & Very often & 5.6\%  \\
       &        &       &        & With more than one type & 22.2\% &            &        &            &        \\ \hline
\end{tabular}
\end{table*}

\subsection{Quantitative Findings}
In this section, we present the key results from CLMM analysis. Detailed results regarding covariates, random effects, model fit, and model formula are provided in~\autoref{apd:model}.

\begin{figure*}[]
    \centering
    \includegraphics[width=.97\textwidth]{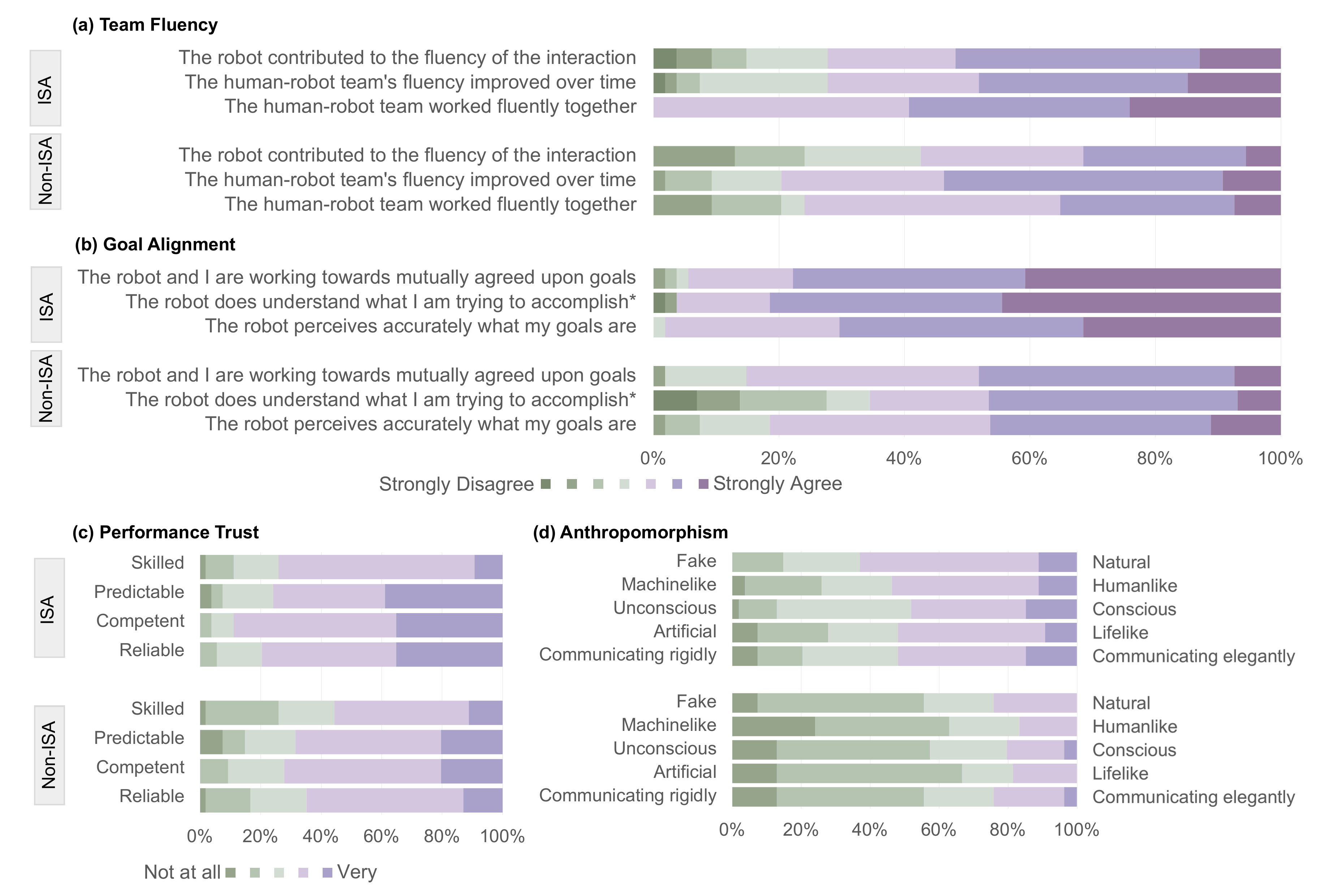}
    \caption{Participant responses on their perceptions of the team fluency (a), goal alignment (b), performance trust (c), and the robot's anthropomorphism (d) under different Speech Modes. 
    (*: This item was originally an inverse item according to~\cite{hoffman2010effects}. To make this figure look consistent, we reversed this item ($current\_score = 8-original\_score$).}
    \label{fig:quan}
\end{figure*}

\subsubsection{RQ1.1: How does a robot's capability to understand indirect speech acts influence the fluency of human-robot teamwork?}

Participants in the ISA group reported significantly greater perceptions of team fluency compared to those in the Non-ISA group ($\beta = 0.961, SE = 0.403, z = 2.382, p =0.017$), as seen in \autoref{tb:result}. Therefore, H1a is confirmed. \autoref{fig:quan}a illustrates the distribution of participants' responses.
The team fluency questionnaire was consistent and reliable (Cronbach's $\alpha = 0.801$)~\cite{hoffman2019evaluating}. 

\subsubsection{RQ1.2: How does a robot's capability to understand indirect speech acts influence the establishment of goal alignment among the human-robot team?}

We observed that the Speech Mode had a significant impact on goal alignment, with participants in the ISA group expressing a stronger belief that they were working toward a mutual goal with the robot ($\beta = 2.309, SE = 0.656, z = 3.518, p < 0.001$). Therefore, H1b is confirmed. This is further illustrated in \autoref{fig:quan}b, which shows their scaled responses.

Moreover, participants' prior experience significantly influenced their perception of goal alignment. Namely, those with greater experience in physical collaborative tasks provided significantly higher scores on the goal alignment scale ($\beta = 0.536, SE = 0.270, z = 1.985, p = 0.047$). The goal alignment questionnaire was also consistent and reliable (Cronbach's $\alpha = 0.794$)~\cite{hoffman2019evaluating}.

\subsubsection{RQ2: How does a robot's capability to understand indirect speech acts influence a human teammate's trust in the robot's performance?}

The ISA group demonstrated significantly higher trust in the robot's performance compared to the Non-ISA group ($\beta = 1.105, SE = 0.493, z = 2.240, p = 0.025$).  Therefore, H2 is confirmed.
The MDMT performance trust questionnaire was consistent and reliable (Cronbach's $\alpha = 0.92$)~\cite{ullman2019mdmt}. Detailed participant responses can be seen in \autoref{fig:quan}c. 

\begin{table*}[]
\caption{The key results from CLMM analysis. Italics are covariates.\vspace{-0.5em}}
\label{tb:result}
\resizebox{0.85\textwidth}{!}{
\begin{tabular}{lllllll}
\hline
 & Fixed Effects & Estimates & Std Error & 95\% CI & z & p-value \\ \hline
Team Fluency & Speech Mode (Non-ISA) & 0.961 & 0.403 & 0.17 -- 1.751 & 2.382 & 0.017* \\
 & \textit{Robot (No)} & 0.004 & 0.110 & -0.211 -- 0.218 & 0.032 & 0.974 \\
 & \textit{Voice Assistant (Never)} & 0.129 & 0.210 & -0.283 -- 0.541 & 0.615 & 0.538 \\
 & \textit{Physical Collaborative Tasks (Never)} & 0.193 & 0.174 & -0.147 -- 0.533 & 1.112 & 0.266 \\ \hline
Goal Alignment & Speech Mode (Non-ISA) & 2.309 & 0.656 & 1.023 -- 3.596 & 3.518 & \textless{}0.001*** \\
 & \textit{Robot (No)} & 0.120 & 0.170 & -0.214 -- 0.453 & 0.701 & 0.483 \\
 & \textit{Voice Assistant (Never)} & -0.099 & 0.316 & -0.719 -- 0.521 & -0.312 & 0.755 \\
 & \textit{Physical Collaborative Tasks (Never)} & 0.536 & 0.270 & 0.007 -- 1.064 & 1.985 & 0.047* \\ \hline
Performance Trust & Speech Mode (Non-ISA) & 1.105 & 0.493 & 0.138 -- 2.072 & 2.240 & 0.025* \\
 & \textit{Robot (No)} & -0.041 & 0.136 & -0.307 -- 0.226 & -0.298 & 0.766 \\
 & \textit{Voice Assistant (Never)} & 0.231 & 0.248 & -0.255 -- 0.717 & 0.932 & 0.351 \\
 & \textit{Physical Collaborative Tasks (Never)} & 0.400 & 0.211 & -0.014 -- 0.814 & 1.892 & 0.058 \\ \hline
Anthropomorphism & Speech Mode (Non-ISA) & 2.708 & 0.674 & 1.387 -- 4.03 & 4.016 & \textless{}0.001*** \\
 & \textit{Robot (No)} & -0.168 & 0.184 & -0.528 -- 0.192 & -0.915 & 0.360 \\
 & \textit{Voice Assistant (Never)} & 0.031 & 0.340 & -0.635 -- 0.697 & 0.092 & 0.927 \\
 & \textit{Physical Collaborative Tasks (Never)} & -0.111 & 0.288 & -0.676 -- 0.454 & -0.385 & 0.701 \\ \hline
\end{tabular}
}
\end{table*}

\subsubsection{RQ3: How does a robot's capability to understand indirect speech acts influence a human teammate's perception of the robot's anthropomorphism?}

\autoref{fig:quan}d shows participants' responses to perceiving the robot's anthropomorphism under different Speech Modes. 
As shown in \autoref{tb:result}, participants in the ISA condition exhibited significantly higher perceptions of the robot's anthropomorphism compared to those in the Non-ISA condition ($\beta = 2.708, SE = 0.674, z = 4.016,$ $p < 0.001$).  Therefore, H3 is confirmed. The anthropomorphism sub-scale of the GSQ questionnaire was consistent and reliable (Cronbach's $\alpha = 0.94$)~\cite{laban2019working}.

\subsubsection{Summary}

Overall, a robot's ability to understand indirect speech acts significantly influences human perception of teamwork fluency, goal alignment, performance trust, and robot anthropomorphism. Regarding the covariates, only participants' prior experience with physical collaborative tasks has a significant positive influence on the human-robot team's goal alignment. Moreover, the Speech Mode has a higher effect on goal alignment and anthropomorphism, followed by a medium effect on performance trust and team fluency.

\subsection{Qualitative Findings}
Two researchers thoroughly analysed participants' 502-minute semi-structured interview recordings~\cite{braun2012thematic}. Additionally, observations made by the experimenters during the experiment that were relevant to the interview findings were also analysed. In the following sections, we present the themes derived from interview responses.

\subsubsection{Reasons for using (In)direct requests}\label{sec:reason}
The most frequently mentioned reason for using indirect requests during collaboration was politeness. Participants preferred to use \textit{``Can you ...?''}—the conventionalised ISA—to show politeness in requests. They believed this approach followed social norms and felt more natural and comfortable. $P4_{Non-ISA}$ also mentioned that using ISA could offer the teammate the option to reject the request. However, $P3_{Non-ISA}$ disagreed, believing it unnecessary to be polite to a robot, which saved their effort. Therefore, she preferred to use direct commands when interacting with robots. Additionally, participants from the Non-ISA group highlighted that even after realising the robot could not understand their indirect requests, they sometimes inadvertently used ISA because it was natural and subconscious.

Due to the subconscious nature, participants who converted indirect requests to direct ones noted using ISA caused less cognitive workload. $P4_{Non-ISA}$ and $P6_{ISA}$ analogised this conversion process to constructing prompts, which requires additional time and effort. However, it was unnatural and more challenging to formulate prompts mentally and articulate them verbally when facing a physical entity, whether it was a robot or a human teammate. Furthermore, $P4_{Non-ISA}$ emphasised that if there were multiple human teammates and a robot teammate in one group, it added unnecessary mental load to switch between direct and indirect communication.

\begin{quote}
    $P4_{Non-ISA}$: ``[On using ChatGPT] It didn't feel like talking to an actual figure. It did make you feel a bit more like I'm not even asking you; I'm just telling you to do something, which just didn't seem natural to me in speech form. But if it was typed out, it would be a bit easier to do that. But I have to say I have to process it a little bit more. If it's just a text screen, then I feel like there's less of a need to express any of that [ISAs] through text form or if it's just like a virtual assistant.''
\end{quote}

Moreover, participants' expectation of the robot's capability influenced their communication strategies. Those who believed the robot had a high level of understanding were more likely to use indirect commands ($P1_{ISA}$, $P15_{ISA}$). Conversely, $P27_{ISA}$ predominantly used direct requests, despite being in the ISA group, as his extensive experience with LLMs led him to doubt the AI's ability to understand implicit requests. He believed direct commands were crucial for successful task completion, even though this approach required more effort to construct explicit commands mentally. Interestingly, participants had differing perceptions regarding the simplicity of commands. $P2_{ISA}$ believed that direct commands were simpler for both humans and robots. In contrast, participants stated that using indirect requests felt simpler and intuitive because it was speaking aloud what was already in mind. \textit{``Like an extension [of mind]''}, said $P13_{ISA}$.

However, there were several differing opinions on direct requests. The most frequently mentioned reason for using direct commands was clarity. Participants believed that direct commands were better suited for tasks requiring precise and nuanced descriptions, whereas indirect requests were more likely to cause ambiguity and misunderstandings. $P5_{Non-ISA}$ and $P17_{Non-ISA}$ further explained that communication strategies exhibit task dependency. For high-risk tasks, direct commands are preferred because unambiguous instructions are critical. In contrast, more complex but low-risk tasks, as well as those requiring intensive collaboration, benefit from indirect and natural communication.

\subsubsection{Adaptation in team fluency}
Participants in the ISA group believed the robot's ability to understand ISA contributed to a higher team fluency. $P1_{ISA}$ and $P18_{ISA}$ agreed indirect commands enabled more flexibility in communication. However, participants in the Non-ISA group reported feeling halted, but most of them further added that it wouldn't be a problem once they adapted. Observations by the experimenters revealed that participants in the Non-ISA group often did not immediately recognise that the issue was due to the misunderstanding of their intentions. Instead, they believed it was a voice recognition problem. As a result, they tended to repeat their indirect requests slowly and word by word, which repeatedly interrupted the collaborative process. Over time, once participants in the Non-ISA group understood that the robot only responded to direct requests, they began using direct commands more consistently, although they occasionally reverted to indirect requests due to the subconsciousness, as discussed in section \ref{sec:reason}. 

According to $P4_{Non-ISA}$, $P8_{Non-ISA}$, and $P11_{Non-ISA}$, although direct requests caused more mental work and initially affected team fluency, they believed this issue would diminish once humans adapted to the robot's communication abilities. \textit{``I think once you get used to it [direct requests], not so much [affects on fluency]. When I realise I have to say things in a certain way, I think it's fine. But at the start, yeah, it's a little bit off.''} said $P8_{Non-ISA}$.
They stated that they had no expectation for the robot to adapt to human communication style, as it was human's responsibility to ensure the robot could understand their instructions. However, not all the users were able to adapt. Despite being aware of the robot’s limitations, $P16_{Non-ISA}$ still preferred to use indirect commands.

\subsubsection{Grounding and goal alignment}
Indirect commands allowed for more flexibility and complexity, fostering a deeper sense of partnership and shared goals ($P9_{Non-ISA}$, $P15_{ISA}$). $P18_{ISA}$ emphasised that a common understanding of the task and goal improved seamless coordination and reduced the likelihood of misunderstandings. According to $P8_{Non-ISA}$ and $P15_{ISA}$, the robot's ability to understand and act on implicit knowledge, similar to human common sense, was crucial for human teammates. This ability included grasping context-specific cues, such as spatial description, incomplete information, and shortened sentences, without requiring detailed explanation. For instance, in the polishing task, $P7_{ISA}$ gave a shortened indirect request \textit{``And the last one please''} to indicate that the robot should turn the hexagonal prism to the final surface, based on the prior context, \textit{``please turn so a different surface is facing me''}. Similarly, in the assembly task, $P13_{ISA}$ used an indirect request with ambiguous information, \textit{``Oh, actually, wait''}, to signal the robot to stop its current movement, with prior context information \textit{``Next, this blue block here needs to go up here on the red block''} and follow-up information \textit{``This red block here needs to go in the middle of this red block here''}. $P5_{Non-ISA}$ explained that he expected the robot to develop a shared understanding based on the context of his commands. As a result, he gave indirect commands, but the robot failed to interpret them correctly. These non-conventionalised ISAs were context-dependent.
Our observations, as well as follow-up interview questions, revealed that participants tended to use non-conventionalised indirect commands when they were confident their intentions were aligned with the robot’s understanding.

\subsubsection{Enhanced performance trust}
Participants in the ISA group reported a high level of performance trust in the interview, which aligns with our quantitative findings. $P1_{ISA}$, $P2_{ISA}$, and $P15_{ISA}$ agreed that they perceived the robot as more capable and reliable once they recognised its ability to understand indirect commands. Some participants in the Non-ISA group believed that the capability and reliability were contingent solely on the robot's task performance rather than its communication abilities. As $P8_{Non-ISA}$ remarked, \textit{``I think as long as there is something that I can say that will make the robot do the task, then it's still capable and reliable''}. 

Additionally, $P7_{ISA}$ emphasised that perceiving the robot as more human-like could raise expectations regarding its performance and lead to potential frustration if errors occurred. Conversely, when the robot was perceived as less human-like, errors were deemed more acceptable. However, some participants also believed that applying social norms, like politeness, in their interaction humanised the robot, which sometimes led to more lenient attitudes towards errors, similar to their reactions to human teammates' mistakes. 

\subsubsection{Perceptions of anthropomorphism}
In line with the quantitative results, the majority of participants acknowledged the robot's ability to understand ISA affected their perception of the robot's anthropomorphism. They agreed that the feeling of human likeness manifested from the ability to understand, even though the voice and tone were still machine-like. $P6_{ISA}$ reported \textit{``I think its ability to understand the implicit language made me feel like somebody was listening''}. However, some participants indicated that additional factors influenced their perceptions. $P12_{Non-ISA}$ highlighted she would \textit{``make small talk with a human, but the robot doesn't.''} Furthermore, $P25_{ISA}$ believed that the sentences used by the robot felt mechanical, which diminished his perception of the robot's human likeness.

Some participants mentioned a feeling of collaboration or control during the experiment. Participants in the Non-ISA group perceived the robot more as a tool or machine rather than a teammate ($P3_{Non-ISA}$, $P5_{Non-ISA}$, $P8_{Non-ISA}$). Participants associated this communication style with a more controlled, mechanical interaction, where the robot was seen as executing specific directives rather than participating in a collaborative process. $P8_{Non-ISA}$ and $P10_{ISA}$ agreed that direct commands required detailed instructions, which reinforced the perception of the robot as a tool needing explicit directions. This approach minimised ambiguity while simultaneously limiting the sense of shared responsibility or joint effort in the task. In this context, the robot was viewed as an extension of the user's will, carrying out predefined actions without critical decision-making. On the contrary, some participants believed the robot's ability to interpret indirect commands indicated a higher level of cognitive processing, similar to human teammates' ability to infer meaning and anticipate actions based on incomplete information. Unlike direct commands, participants perceived the robot more as a teammate rather than a tool when using indirect commands. This perception resulted from the robot's ability to understand and respond to more nuanced and context-rich communication $P1_{ISA}$, $P2_{ISA}$, $P18_{ISA}$, $P18_{ISA}$). Indirect commands were often used in a more conversational tone, suggesting a partnership where the robot was expected to understand the intent behind the instructions and act accordingly. This contributed to participants' perception of the robot's anthropomorphism ($P10_{ISA}$, $P13_{ISA}$, $P18_{ISA}$).

\begin{quote}
    $P17_{Non-ISA}$: ``I think, because indirect commands and subtext is a very human-feeling thing. So when I'm just giving it a direct command, it feels more like I'm just putting an input into a machine. Whereas with the indirect commands, it feels more like I'm having a conversation with someone.''
\end{quote}

When discussing future usage, participants also expressed concerns about different aspects of the robot's anthropomorphism. Participants believed that they preferred a robot with an extremely human level of understanding but not one that mimicked human tone, voice, or appearance. $P1_{ISA}$ mentioned that it also depends on the type of task, \textit{``If it's a vacuum cleaner, I want it to be less human cause it's just vacuuming the floor, you know. But if we're working on tasks kind of like this [our study], where it would be normal for two humans to work together, then I would definitely want the robot to be more human just so it's easier to communicate and get things done quickly.''}

\subsubsection{Expectations on LLM}
As this study used the Wizard-of-Oz method, participants assumed that the ISA robot was implemented with an LLM. Several participants compared the robot’s capabilities with their prior experiences using voice assistants and commercial LLMs. $P25_{ISA}$, $P32_{Non-ISA}$, and $P36_{Non-ISA}$ believed that a novel LLM should be capable of handling indirect requests, at least the conventionalised ones, i.e. \textit{``Can you ...?''.} However, $P25_{ISA}$ also acknowledged that he tended to be more direct when interacting with a text-based LLM. $P32_{Non-ISA}$ believed that using more indirect and polite language with ChatGPT usually yielded better outcomes. \textit{``You have to be very patient''}, $P32_{Non-ISA}$ remarked.

In contrast, $P27_{ISA}$ explained that his experience with LLMs led him to doubt the robot’s ability to comprehend indirect commands, which prompted him to provide explicit instructions to ensure task success. As a result, he primarily used direct commands during the collaboration in our study. $P4_{Non-ISA}$ and $P6_{ISA}$ concurred that ChatGPT usually performed better when using direct commands. Despite this, they both used numerous indirect commands in this study, noting that verbal commands differ from written commands as they provide less time to construct the prompts, and the formation of commands is often ad-hoc, leading to more ambiguity and incomplete sentences. Furthermore, $P6_{ISA}$, who expressed doubts about the implementation of LLMs in our robot, raised concerns about their effectiveness in real physical-embodied scenarios, arguing that LLMs would likely struggle in such contexts.

\begin{quote}
    $P6_{ISA}$: ``[When interacting with the robot] I think because I'm referring to things that exist in space as opposed to a concept that exists just in our mind. So if we're talking about something like, What's the difference between a plant cell and, you know, an animal cell? It's got a text-based understanding of that. But because this [interacting with a robot] is referring to a real embodied scenario, I think it would struggle to do anything with this.''
\end{quote}

%% file: sections/60-discussion.tex
\section{Discussion}
In this section, we discuss the impact of indirect speech acts in human-robot collaboration based on our quantitative and qualitative findings. 

\subsection{ISA's subtle role in teamwork}
\subsubsection{Adaptation and synchronisation}
Participants reported that a robot capable of understanding indirect requests made fluent collaboration easier. This is consistent with our quantitative results. Moreover, participants in the Non-ISA group reported adapting to the robot's communication ability to increase team fluency over time. Previous research shows human collaborators tend to have synchronisation on their vocalisation and neural activity when selecting words to convey contextual meanings during conversations~\cite{abney2021cooperation, zada2024shared}. Moreover, literature further suggests that people unconsciously mirror their linguistic structures with their interlocutors, regardless of being a human or computer, which facilitates efficient interactions~\cite{branigan2010linguistic}. However, the robot in the Non-ISA group failed to reciprocate linguistic convergence by adapting to their human collaborators.

Our findings revealed that the absence of robots' ability to interpret ISAs (Non-ISA group) necessitated greater adaptation efforts from participants during collaboration. This adaptation process was reported to be time-consuming and requiring increased cognitive effort. The robot's failure to perform its role as an effective communication partner forced participants to take on the full responsibility of adapting, increasing their effort and disrupting the division of labour~\cite{clark1996using, dale1995computational}. Previous studies have shown that ideal robot teammates should be able to adapt their communication to establish common ground for shared environment~\cite{chai2016collaborative}. \textbf{Therefore, it is essential to enhance robots’ ability to adapt to individuals' communication styles, for instance, using indirect requests.} Besides, to be accessible to diverse populations, robots should adapt to users’ speech styles, considering factors like ``age, gender, dialect, domain expertise, task knowledge, and familiarity with the robot.''~\cite{marge2022spoken}. Our findings support this call, while highlighting the importance of indirect speech. As collaborative robots enter real-world settings, it is suboptimal to expect groups, like children or the elderly, to adapt to the robot's communication style. 

\subsubsection{Grounding}
Previous research in human-robot interaction conducted limited exploration of non-conventionalised ISAs (i.e. context-dependent ISAs), usually focused on the effect of politeness (i.e. conventionalised ISAs)~\cite{seok2022cultural, williams2018thank}. In our study, we discovered significant effects of non-conventionalised ISAs on team grounding. The interview findings provided an explanation for the questionnaire results, which showed that the ISA group had a significantly higher perception of goal alignment compared to the Non-ISA group. Participants who effectively used indirect requests, particularly non-conventionalised ISAs, to communicate with the robot felt more confident in having established a shared understanding with their robot teammate. 
Moreover, participants mentioned that conventionalised ISAs (e.g., ``Can you...?'') offer the teammate an option to reject the request, consistent with Searle’s~\cite{searle1975indirect} theory, which explains that ISAs also contribute to facilitating the exchange of intentions between teammates. 

In contrast to human-human collaboration, the use of ISAs is nuanced by the users' expectations. Some participants, having no expectation of the robot's ability to understand ISAs, opted to use only direct requests, even when interacting with a robot capable of interpreting ISAs. This finding complements the results of~\cite{briggs2017enabling}, which showed that individuals continue using ISAs when interacting with robots that cannot comprehend them—a pattern also observed in our study. This shows that the user's prior expectations, or mental models, of the robot's capabilities play a strong role in people's decision to use or avoid ISAs. It could be important for a robot teammate to explicitly communicate its capabilities to interpret ISAs. At the start of a collaboration, for example, the robot might say \textit{``Please just give me clear, precise, direct instructions''}.

With human-agent teaming on the rise, goal alignment has emerged as a critical yet unresolved challenge~\cite{bhat2024value, zhang2025implicit}. Previous research has focused on  approaches that model goal alignment and assess its effects~\cite{li2022modeling, sanneman2023validating}. Our findings suggest that the successful use of indirect requests in communication can act as an indicator of mutual understanding within the team. Consequently, proactively \textbf{incorporating implicatures into the robot’s verbal communication may be an effective strategy for signalling the robot’s accurate comprehension of the human teammate’s intentions}. However, ISAs can also introduce ambiguities, requiring the robot to more effectively manage dialogue failures and repair mechanisms~\cite{kontogiorgos2021systematic}. The appropriate usage of this strategy not only enhances the explainability of the robot's mental state but also maintains the flow of teamwork without interruptions.

\subsection{ISA's subtle role in trust}
Qualitative findings suggest that the robot's ability to understand ISAs either positively impacted or did not affect participants' trust as long as tasks were completed successfully. This qualitative feedback supports the quantitative results, indicating that understanding ISAs significantly enhances trust, although trust remained high in the Non-ISA group due to successful task execution. Moreover, some participants felt that using indirect requests enhanced their perception of the robot’s anthropomorphism. This result aligned with the findings in~\cite{emnett2024using}, claiming that a robot's speech anthropomorphism should not be limited to tone and voice but also to directness. Previous studies conclude that robots with higher anthropomorphism in appearance (i.e. looking more human-like) may induce higher functionality expectations~\cite{duffy2003anthropomorphism} and trust~\cite{natarajan2020effects}. Our study adds to these findings that higher human-like understanding in verbal communication may induce higher performance trust. 

However, our study represented an ideal scenario where the robot made no mistakes. In real-world settings, execution and communication failures are common. During the interview, some participants suggested that higher performance trust could result in elevated expectations, potentially causing greater frustration when the robot makes errors. Conversely, participants with a higher perception of the robots' partnership believed they would be more forgiving of the robot’s potential mistakes. Similar contradictory findings have been reported in previous research. \citet{salem2013err} observed that robots displaying occasional incorrect gestures were perceived as more likeable than those that performed perfectly. A follow-up study~\cite{salem2015would} found opposing results, but also suggested that the level of anthropomorphism and the severity of the error may influence these differing reactions. In our experiment, we used a robot with anthropomorphic features, including a head, neck, arm, and torso. Some participants noted during the interviews that their perceptions might differ if the robot were less human-like, such as a vacuum robot. \textbf{Given that real-world interactions are more prone to errors, it is crucial to carefully consider the potential negative effects on trust when employing natural and implicit verbal communication in collaboration.} Although there are no widely recognized studies analysing users' speech acts when a robot fails, \citet{kontogiorgos2021systematic} found that humans tend to emphasise vowels and speak more loudly when robots make errors. Future research could explore users' speech directness in response to robot errors, particularly in relation to the level of anthropomorphism, timing and severity of the failure~\cite{rossi2017timing}.

\subsection{Task- and context-dependency}
Contrary to our assumptions, indirect requests are not suitable for all situations. The usage of ISAs is task-dependent. Participants responded that \textbf{ISAs were preferred when collaborating on repetitive, low- to medium-risk tasks, as well as tasks requiring high coordination}. For high-risk tasks, the explicitness of direct requests is safer, as it provides clearer and more precise descriptions of the required actions. Simple and repetitive tasks typically require less verbal communication, and participants often use shortened indirect requests based on the mutual understanding of the task they have built before, such as \textit{``Next''}. In less collaborative tasks, participants prefer fewer, clearer instructions over back-and-forth dialogue, prioritising efficiency and precision over an intuitive and low-effort interaction experience during task completion.

\textbf{The use of indirect requests is also highly context dependent.} Unlike written commands given to virtual AI assistants, verbal commands are given less time to formulate and are often subconsciously phrased as indirect requests during physical collaboration. Furthermore, since collaborative tasks typically involve continuous interaction and sequential sub-tasks, indirect requests often rely on prior commands and actions. Interpreting these requests requires the ability to reference previous interactions that are related. Researchers suggested ISAs are less semantically related to their immediate context than direct speech acts; however, they gain relevance when interpreted correctly in light of broader context ~\cite{boux2023cognitive}. Previous research highlights the substantial impact of incorporating task context on improving the prediction of ISA usage, emphasising the need for models that account for contextual and intentional factors~\cite{smith2022leveraging}. In HRC, which involves real-world interactions, gestures are frequently employed, further facilitating the use of indirect commands. Additionally, implicatures often rely on real-world information, such as the location of objects. A previous study explored how locative expressions embedded in indirect commands are interpreted~\cite{lamm2017pragmatics}. The physical affordances of the environment, which embed rich semantic information~\cite{gao2024physically, brohan2023can}, can readily prompt the use of implicatures in indirect requests. Future research could focus on developing solutions that address broader conversational contexts and link physical environments to improve the accuracy of robots' interpretation of non-conventionalised indirect requests, where large language and vision models have shown strong potential due to their long-term context-sensitive attention and multi-modal reasoning capabilities~\cite{zhong2024memorybank, sermanet2024robovqa}.
\vspace{-0.5em}

\subsection{Limitations and future work}
With the rapid development of LLMs and enhanced reliability and affordability of robot hardware, the use of natural language as an interface for daily human-robot collaboration is becoming increasingly feasible. However, some participants noted that LLMs performance in interpreting indirect requests, based on their experiences with commercial models, varied significantly, indicating that while LLMs show some capacity for interpreting implicature, their reliability remains inconsistent. This challenge has also been highlighted and explored by other researchers~\cite{ruis2024goldilocks, jin2024reasoning}. Moreover, the context- and task-dependent nature of using ISAs in physical human-robot collaboration presents additional challenges, particularly in integrating visual and physical information. Therefore, a future evaluation is necessary before deploying LLMs in commercial collaborative robots, along with the development of specialised datasets and fine-tuning techniques. There is also potential for expanding the scope to broader conversational contexts and linking physical environments to enhance robots' interpretation of non-conventionalised indirect requests, where recent advances in language and vision models offer promising solutions with their context-sensitive attention and multi-modal reasoning capabilities. 

There are several limitations in our study. First, although participants represented a wide age range, most were recruited from a university campus, with the majority being college students, which limits the generalisability of our findings to the broader population or specific demographic groups. Second, due to the use of teleoperation to ensure safety, participants noted that the robot’s arm movements were slow and unsteady, which may have influenced their perception of the robot’s capabilities. Third, this experiment employed a Wizard-of-Oz setup, which created an error-free scenario, allowing us to focus on analyzing users' behaviour in using ISAs and comparing across Speech Modes. However, robot errors are unavoidable in real-world applications. Future research should investigate how robot errors influence the directness of user speech, as well as their impact on collaboration performance and experience. Fourth, this experiment did not control the amount of ISAs each participant used during their interaction session to maintain natural interactions. The effect of this variation was considered in a broader sense by accounting for participants and task type as random effects rather than precisely measuring the number of ISAs used. Additionally, this study exclusively examined the impact of robots' ability to understand ISAs. Future research should explore the effects of robots' ability to generate ISAs in collaborative settings. Another important area for investigation is the influence of robots using ISAs on human communication patterns, particularly whether a robot’s use of ISAs encourages humans to reciprocate with indirect speech. This dynamic could impact users' tolerance for dialogue errors, with effective ISA use potentially fostering greater flexibility and tolerance for minor mistakes, while improper handling of ISAs could reduce trust and interaction fluency.
\vspace{-0.5em}

%% file: sections/70-conclusion.tex
\section{Conclusion}
In this study, we investigated the impacts of indirect speech acts on human-robot collaboration. Our findings highlight that the robot's ability to interpret ISAs plays a crucial role in verbal communication, though the implications of this ability vary depending on context and task. Our results suggest that ISAs hold significant potential as a communication tool to facilitate team fluency, goal alignment, and trust in HRC when applied appropriately. Robots with the ability to understand indirect requests can also increase human perception of anthropomorphism, which enhances the sense of partnership and results in a better collaborative experience. We further explored the human motivations for using indirect requests and the underlying factors driving these impacts using qualitative analysis.  

Future research should focus on assessing language models' ability to interpret implicatures in indirect requests, provide appropriate ISAs, and develop large language models capable of nuanced, context-aware interactions for robotic systems. Moreover, given the inherent ambiguity of ISAs, designing effective backchanneling mechanisms to prevent misunderstandings and convey uncertainty is equally important. We advocate for careful integration of both direct and indirect verbal communication into the design and evaluation of collaborative robots, ensuring that ISAs are neither overlooked nor overused in inappropriate contexts.